\documentclass[prl,twocolumn,superscriptaddress]{revtex4-2}
\usepackage{physics}
\usepackage{mathtools}
\usepackage{braket}
\usepackage{amsfonts}
\usepackage{amssymb}
\usepackage{amsmath}
\usepackage{graphicx}
\usepackage{dcolumn}
\usepackage{bm}
\usepackage{units}
\usepackage{multirow}
\usepackage{CJKutf8}
\usepackage{subfigure}
\usepackage{color}
\usepackage{url}
\usepackage{ulem}
\usepackage[colorlinks,linkcolor=blue,anchorcolor=blue,citecolor=blue,urlcolor=blue]{hyperref}
\usepackage{color,soul}
\usepackage{titlesec}
\usepackage{cleveref}

\titleformat{\section}[runin]
{\normalfont\itshape}
{\S\ \thesection.}{.5em}{}[.--- ]
\titlespacing{\section}
{\parindent}{1.5ex plus .1ex minus .2ex}{5pt}

\begin{document}
\title{Spin wave amplification through superradiance}

\author{X. R. Wang}
\email[Corresponding author: ]{phxwan@ust.hk}
\affiliation{Physics Department, The Hong Kong University of Science and 
Technology, Clear Water Bay, Kowloon, Hong Kong}
\affiliation{HKUST Shenzhen Research Institute, Shenzhen 518057, China}
\affiliation{William Mong Institute of Nano Science and Technology, 
The Hong Kong University of Science and Technology,
Clear Water Bay, Kowloon, Hong Kong, China}

\author{X. Gong}
\affiliation{Physics Department, The Hong Kong University of Science and 
Technology, Clear Water Bay, Kowloon, Hong Kong}

\author{K. Y. Jing}
\affiliation{Physics Department, The Hong Kong University of Science and 
Technology, Clear Water Bay, Kowloon, Hong Kong}

\date{\today}

\begin{abstract}
Superradiance is a phenomenon of multiple facets that occurs 
in classical and quantum physics under extreme conditions. 
Here we present its manifestation in spin waves under 
an easily realized condition. We show that an interface 
between a current-free (normal) ferromagnetic (FM) region 
and a current-flow (pumped) FM region can be a spin wave 
super-mirror whose reflection coefficient is larger than 1. 
The super-reflection is the consequence of current-induced 
spectrum inversion where phase and group velocities of 
spin waves are in the opposite directions. 
An incident spin wave activates a backward propagating 
refractive wave inside pumped FM region. The refractive spin 
wave re-enters the normal FM region to constructively 
interfere with the reflective wave. It appears that the pumped 
FM region coherently emits reflective waves, leading to a 
super-reflection. The process resembles superradiance of a 
spinning black hole through the Hawking radiation process, 
or Dicke superradiance of cavity photons inside population 
inverted media.
\end{abstract}
\maketitle

Spin waves have attracted much attention in spintronics due to 
their rich physics \cite{review1,Xiansi2012,Pengyan2011} and 
potential applications 
\cite{Xiansi2018,Chumak2014,Csaba2017,Barman2021}. 
One challenge in spin-wave based applications is the strong spin 
wave damping such that it can only propagate a relatively short 
distance, order of millimetres even for yttrium iron garnets 
\cite{Gerrit2020,Liu2018} with the lowest damping coefficient. 
Thus, an effective method for amplifying an arbitrary spin 
wave is demanded in order to realize the power of spin waves. 

Many efforts have been made on spin wave amplification. 
The anti-damping nature of spin transfer torques (STTs) 
\cite{Padron2011,Seo2009,An2014,Woo2017} and spin orbit 
torques \cite{Divinskiy2018, Demidov2020,Nikitchenko2021} 
was considered first. The Landau-Lifshitz-Gilbert (LLG) 
equation suggests that spin torques can exponentially 
amplify all spin waves \cite{Seo2009}. Unfortunately, 
experiments and micromagnetic simulations observed only the increase of spin wave propagation distance, but no spin wave
amplification \cite{Seo2009,Padron2011,An2014,Divinskiy2018,
Woo2017,Demidov2020,Nikitchenko2021} with probably one 
exception \cite{Merbouche2023}. 
Parametric pumping has also been used to amplify spin waves. 
The idea is to resonate spin waves with external oscillating 
fields such as an electromagnetic or a mechanic wave \cite
{Kolodin1998,Verba2014,Hillebrands2017,Pirro2022,Chowdhury2015,
Graczykappa2017}. At resonance, energy from external fields 
can be transformed into the energy of a spin wave. 
However, parametric pumping requires specific resonant conditions 
that are only applicable to a particular wave frequency.
The Klein paradox \cite{klein,book}
has been proposed to amplify spin waves. The concept involves 
using STTs or spin-orbit torques to create a spin wave 
analogue of black and white holes \cite{Duine2019,Harms2022} such 
that positive-energy and negative-energy spin wave modes 
can be coupled together in an inhomogeneous FM film. 
Experimental realization of this proposal is yet to be done.
Nonlinear 3-magnon and 4-magnon process have also been proposed to 
amplify spin waves and to create spin wave combs \cite{Pen2022}. 
This approach often requires stringent conditions and for a specific mode.

Reflection and refraction are general wave phenomena occurring at 
an interface between two homogeneous media. A general rule is that 
intensities of both reflective and refractive waves are smaller 
than that of the incoming wave due to the energy conservation law. 
The frequency (energy) minimum of wave spectrum in a normal 
media is at zero wave number such that the group and phase 
velocities are in the same directions (for isotropic media). 
An adiabatic spin-transfer torque from an electric current along the spin wave propagating direction can invert the spin wave 
spectrum in which the group and phase velocities in certain frequency window could be in the opposite directions. Such a current-flow region is termed a pumped medium. In an electric-current-flow magnetic film, spin wave spectrum is 
inverted along the current direction. Surprisingly and 
interestingly, the reflection coefficient of a wave at the interface between a normal medium and a pumped medium can be greater than 1. It appears that pumped region is coherently emitting spin waves to the normal region. The process resembles superradiance \cite{book} of a spinning black hole through the Hawking radiation, or the Dicke superradiance of cavity photons inside population inverted media. 

In this letter, we show that an interface between a current-free 
and a current-flow FM regions can be super-mirror for spin 
wave in a frequency range tunable by an electric current. 
A wave packet in the pumped medium propagates forward 
(positive group velocity) while waves of given negative 
wave numbers move backward (negative phase velocities) to 
emit backward spin waves at the interface. The emitted waves 
interfere constructively with the reflective waves, leading to 
the super-reflection. The physics is clearly revealed in a 
generic FM model and is confirmed by micromagnetic simulations.

\begin{figure}[h]
	\centering
	\includegraphics[width=8.5cm]{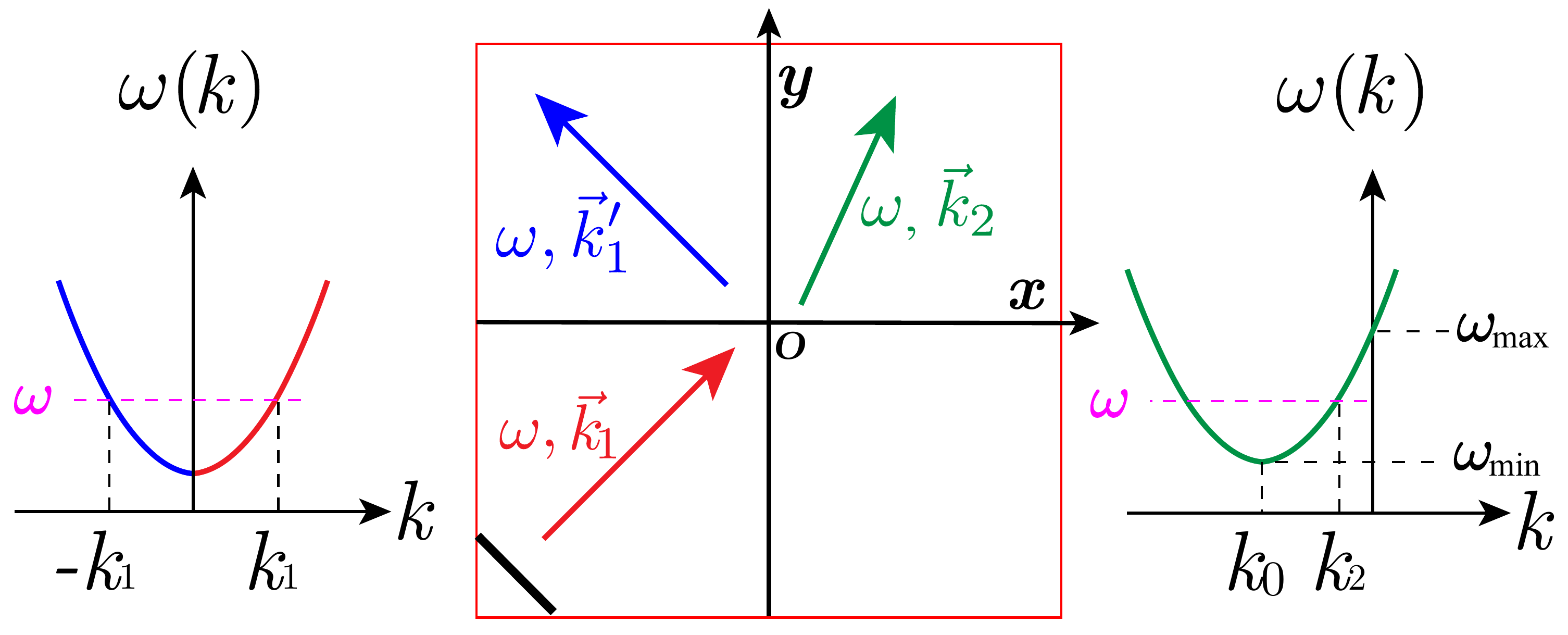} 
\caption{Schematic illustration of a incident spin wave of 
frequency $\omega$ and wave vector of $\vec{k}_1$, reflective 
wave of $\omega$ and $\vec{k}'_1$, and refractive wave of 
$\omega$ and $\vec{k}_2$. The black bar donates spin wave 
source from an oscillating magnetic field. 
The normal and inverted spectra are schematically illustrated 
in the left and right panels, respectively. }
\label{model}
\end{figure}

We consider a thin magnetic film of thickness $d$ laying in the 
$xy$-plane as shown in Fig. \ref{model}. The film on the left 
($x<0$) is current-free and film on the right ($x>0$) has an 
electric current density $\vec j$ flowing along the 
$x$-direction ($\hat x$) such that the spin wave spectrum 
minimum is at wave vector $\vec k=(k_0,0)$ ($k_0\propto j<0$). 
The middle panel of Fig. \ref{model} illustrates an incident 
wave of $\omega$ and $\vec{k}_1$, and reflective wave of $\omega$ 
and $\vec{k}'_1$ and refractive wave of $\omega$ and $\vec{k}_2$. 
The left and right panels are the schematic plots of normal and 
inverted spin wave spectrum along $\vec k=(k,0)$, respectively. 
$\omega$ is minimum at $k_0\ne 0$ in the inverted spectrum and 
phase and group velocities are opposite in sign for $\omega\in[
\omega_{\mathrm{min}}, \omega_{\mathrm{max}}]$. Magnetization 
state $\vec M(\vec x,t)=M_s\vec m(\vec x, t)$ with saturation 
magnetization $M_s$ and direction $\vec m$ is described by the 
magnetic energy, 
\begin{equation}
E=d\int [A(\nabla\vec{m})^2 +K_u(1-m_z^2)-\frac{1}{2}\mu_0 
M_s\vec H_d\cdot\vec{m} ]\mathrm{d}x\mathrm{d}y,
\end{equation}
where $A$, $K_u$, $\vec H_d$ and $\mu_0$ are the exchange 
stiffness constant, the crystalline magnetic anisotropy, the 
demagnetizing field and the vacuum permeability, respectively. 
The demagnetization effect can be included in the effective 
anisotropy $K=K_u-\mu_0M_s^2/2$ for an ultra-thin film \cite{paper4}. Her we mimic Co/Pt film with 
$M_s=\mathrm{0.6\:MA\:m^{-1}}$, $A=\mathrm{10\:pJ\:m^{-1}}$, 
$K_1=\mathrm{0.6\:MA\:m^{-3}}$ in normal region and 
$K_2=\mathrm{0.9\:MA\:m^{-3}}$ in the inverted region. 
Magnetization dynamics is governed by the LLG equation,
\begin{equation}\label{LLG}
\frac{\partial \vec{m}}{\partial t} = -\gamma  \vec{m} \times 
\vec{H}_{\rm eff} + \alpha \vec{m}\times 
\frac{\partial \vec{m}}{\partial t} - (\vec{u}\cdot \nabla)\vec{m},
\end{equation}
where $\gamma$ and $\alpha$ are respectively the gyromagnetic 
ratio and the Gilbert damping constant. 
$\alpha=0.001\sim0.005$ is used in this study. 
The effective field is $\vec{H}_{\rm eff}=-(\mu_0 M_s )^{-1}
\delta E/ \delta\vec{m}$. $\vec{u}=\frac{P\mu_B}{e M_s}\vec{j}$ 
is the spin drift velocity, here $P$, $\mu_B$, $e(>0)$, and 
$\vec{j}$ are the spin polarization, the Bohr magneton, 
electron charge, and the applied current density, respectively.
$\vec{j}= \mathrm{10^{13}\hat x \:A\:m^{-2}}$ \cite{Tsoi1998,Ji2003} 
in the inverted medium is used throughout this study with spin polarization $P=1$ if not stated otherwise. 
Although only an adiabatic STT is included here, including a 
non-adiabatic STT will not change physics, see evidences and discussions in Supplemental Material. 

Eq. \eqref{LLG} is numerically solved by MuMax3 \cite{mumax3}, 
and spin wave spectrum is obtained by the Fourier analysis of thermally generated random $\vec m(\vec x, t)$ directly from 
MuMax3 (see Supplemental Material for details). To investigate 
spin wave reflection and refraction, we simulate Eq. \eqref{LLG} 
with an oscillating magnetic field along a narrow strip at 
lower left corner of the film, the black bar in Fig. \ref{model}.
To avoid boundary effects, we add absorb regions at boundaries 
with a large Gilbert damping $\alpha = 0.5$. 

\begin{figure}[h] 
   \centering
   \includegraphics[width=8.5cm]{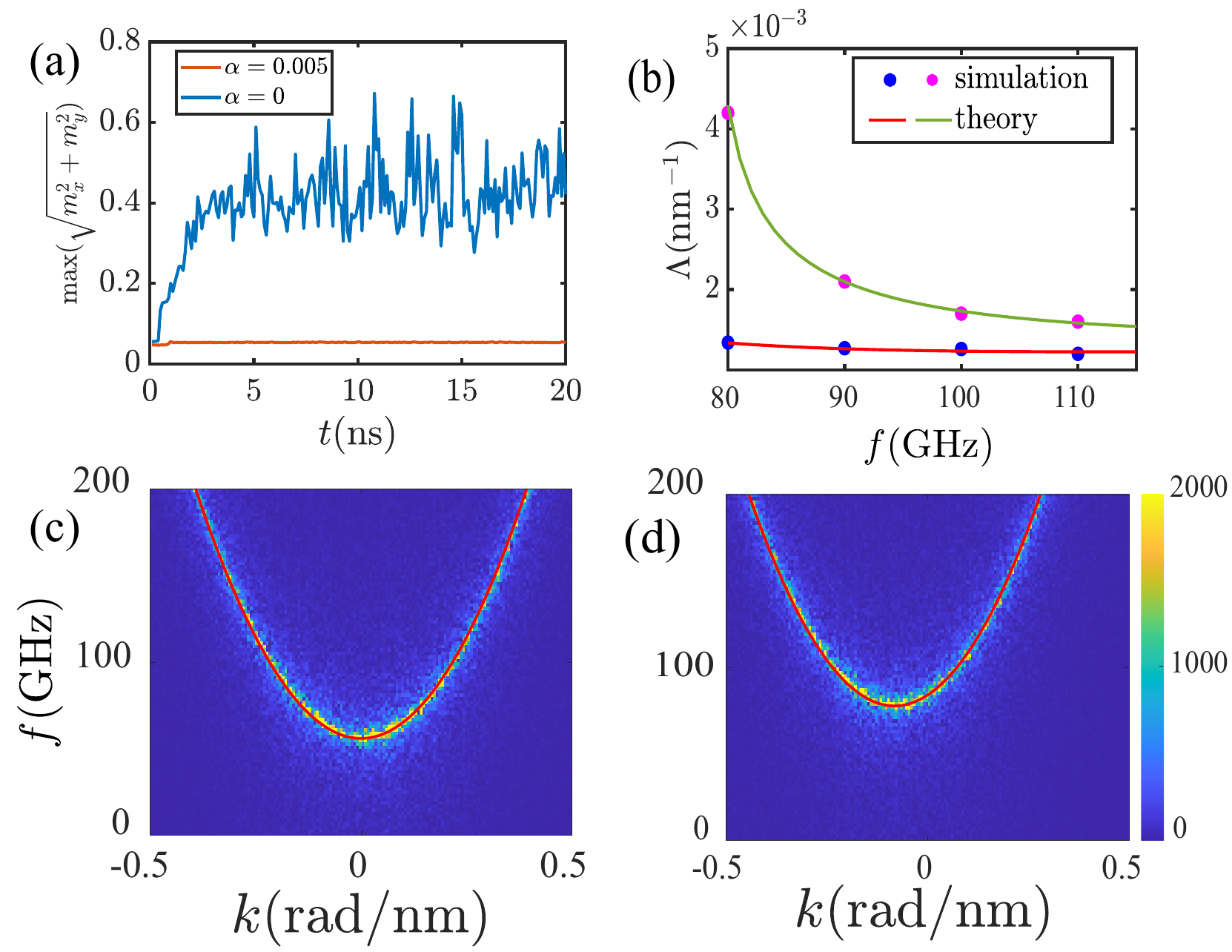} 
\caption{(a) $\vec{m}=\hat{z}$ under a current is no longer 
an equilibrium state when $\alpha=0$ (the blue curve), but  
remains an equilibrium state with a tiny $\alpha=0.005$. 
(b) $\Lambda$ vs $f$ for $\alpha=0.005$ in region 1 (blue 
dots and the red line) and 2 (purple dots and the green line).
Density plot of the Fourier transformation of $\varphi(x, t)$ 
in $f k$-plane shows spin waves spectrum in current-free 
(c) and current-flow (d) regions. }
   \label{spectrum}
\end{figure}

Without STTs, the LLG equation describes dissipative dynamics \cite
{xrw1} if $\alpha\ne 0$ and the system ends at a ferromagnetic 
state, $\vec{m}=\pm \hat{z}$. The LLG equation is a Hamiltonian 
system if $\alpha=0$ and the system shall move on equal-energy 
contours \cite{xrw1}. In the presence of STT and with a damping, 
$\vec{m}=\pm \hat{z}$ are stable against STT agitations due to the 
quick damping of spontaneously generated spin waves. Without a 
damping and with a STT, the STT can do work in an inhomogeneous region 
(inevitably near sample boundary and around defects in reality) 
and generates all types of spin waves such that no static state 
is possible as shown by the blue curve in Fig. \ref{spectrum}(a). 
Spin waves are the normal modes of small fluctuation of $\vec{m}$, 
$\vec m=m_x\hat{x}+m_y\hat{y}+\hat{z}$, $m_x,m_y\ll 1$, govern by 
\begin{equation}\label{matrixeq}
\left( \begin{matrix}
\partial_t + \vec{u}\cdot\nabla& -A'\nabla^2+K'+\alpha\partial_t\\
A'\nabla^2-K'-\alpha\partial_t & \partial_t + \vec{u}\cdot\nabla
\end{matrix} \right)\left( \begin{matrix} m_x  \\
m_y \end{matrix} \right) =0,
\end{equation}
where $A'=\frac{2\gamma A}{\mu_0 M_s}$, $K'=\frac{2\gamma K}{\mu
_0M_s}$ and $\vec{u}=\frac{P\mu_B}{eM_s}\vec{j}$ (0 for $x<0$). 
In terms of $\varphi\equiv m_x+im_y$, Eq. \eqref{matrixeq} becomes, 
\begin{equation}
-i\frac{\partial \varphi}{\partial t}=\frac{1+i\alpha}{1 
+\alpha^2} (-A'\nabla^2 + K'+i\vec{u}\cdot \nabla) \varphi ,
\end{equation}
with spin wave solutions of   
$\varphi = \sum_{k}a_k e^{-\vec{\Lambda}\cdot\vec{r}} 
e^{i(\omega t -\vec{k}\cdot\vec{r})}$  
and spectrum of 
\begin{equation}
\omega  = A' |\vec{k}|^2 + K' +\vec{u}\cdot \vec{k}.
\label{spectrumeq}
\end{equation}
$a_k e^{-\vec{\Lambda}\cdot\vec{r}}$ is the exponential decay 
of the amplitude. The decay length $|\vec{\Lambda}|^{-1}$ 
is given by $\vec{\Lambda}\cdot\vec{v} = \alpha\omega$.
$\vec{v} =2A' \vec{k}$ ($2A' \vec{k} + \vec{u}$) is the 
group velocity in the current-free (current-flow) region.  
Spin waves are gapped with gap $K'_1$ in region 1, and 
$K'_2 -u^2/4A'$ in region 2. Figures \ref{spectrum}(c) and 
(d) show the spectrum of spin waves along $\vec{k}=(k, 0)$,  
where colors denote density plot and curves are analytic 
formula Eq. \eqref{spectrumeq} (in terms of linear frequency $f={\omega}/{2\pi}$), which agrees with simulations.  

\begin{figure}[h] 
   \centering
   \includegraphics[width=8.5cm]{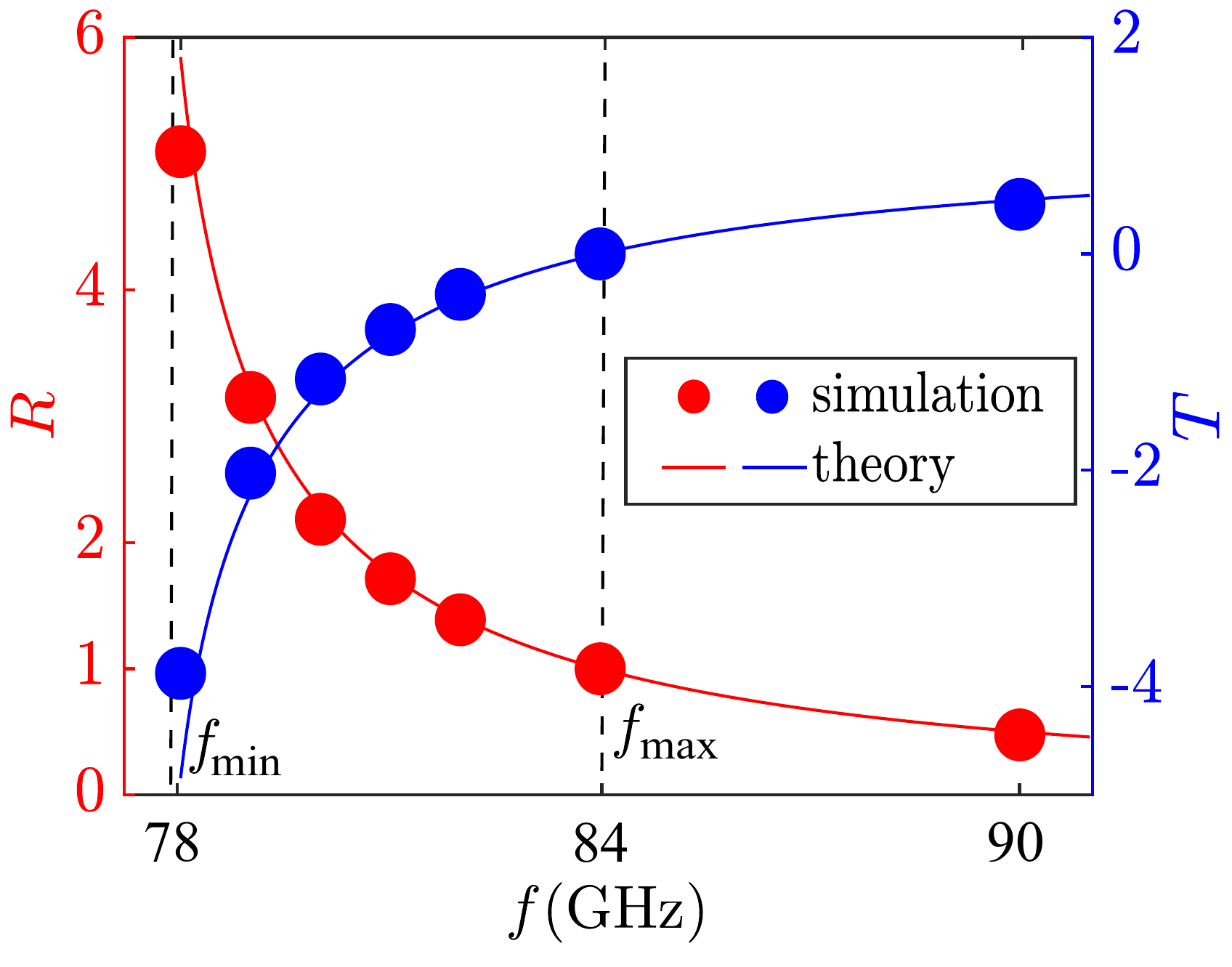} 
\caption{Reflection and transmission coefficients as functions 
of $f$ from simulations (red and blue dots) and 
theory (the red and blue line). Two dashed lines mark 
$f_{\mathrm{min}}$ and $f_{\mathrm{max}}$, respectively.}
   \label{reflection}
\end{figure}

Spin wave satisfies continuity conditions at the interface, 
$\varphi (x=0^-)=\varphi (x=0^+)$ and $\nabla\varphi(x=0^-) 
= \nabla\varphi(x=0^+)$. For a normally incident spin wave of 
frequency $\omega$, $\varphi (x,t)=B e^{-\Lambda_1 x}e^{i
(\omega t-k_1x)}+Ce^{\Lambda_1 x} e^{i(\omega t+k_1 x)}$ 
for $x<0$ and $\varphi (x,t)=De^{-\Lambda_2 x} e^{i(\omega 
t-k_2 x)}$ for $x>0$. Reflection coefficient $R$ and 
transmission coefficient $T$ can easily be obtained, 
\begin{equation}\label{reflectioneq}
\begin{aligned}
R=\frac{|C|^2}{|B|^2}=\left (\frac{k_1-k_2}{k_1+k_2}\right )^2 \\
T=\frac{k_2 |D|^2}{k_1 |B|^2}=\frac{4 k_1 k_2}{(k_1+k_2)^2}.
\end{aligned}
\end{equation}
If $\omega$ is in the inverted spectrum, $k_2<0$, $R>1$ and 
negative $T<0$ show the backward propagation.   
This super-reflection is confirmed from the micromagnetic 
simulations as shown in Fig. \ref{reflection} for $k_2<0$. 
\begin{figure}[h] 
   \centering
   \includegraphics[width=8.5cm]{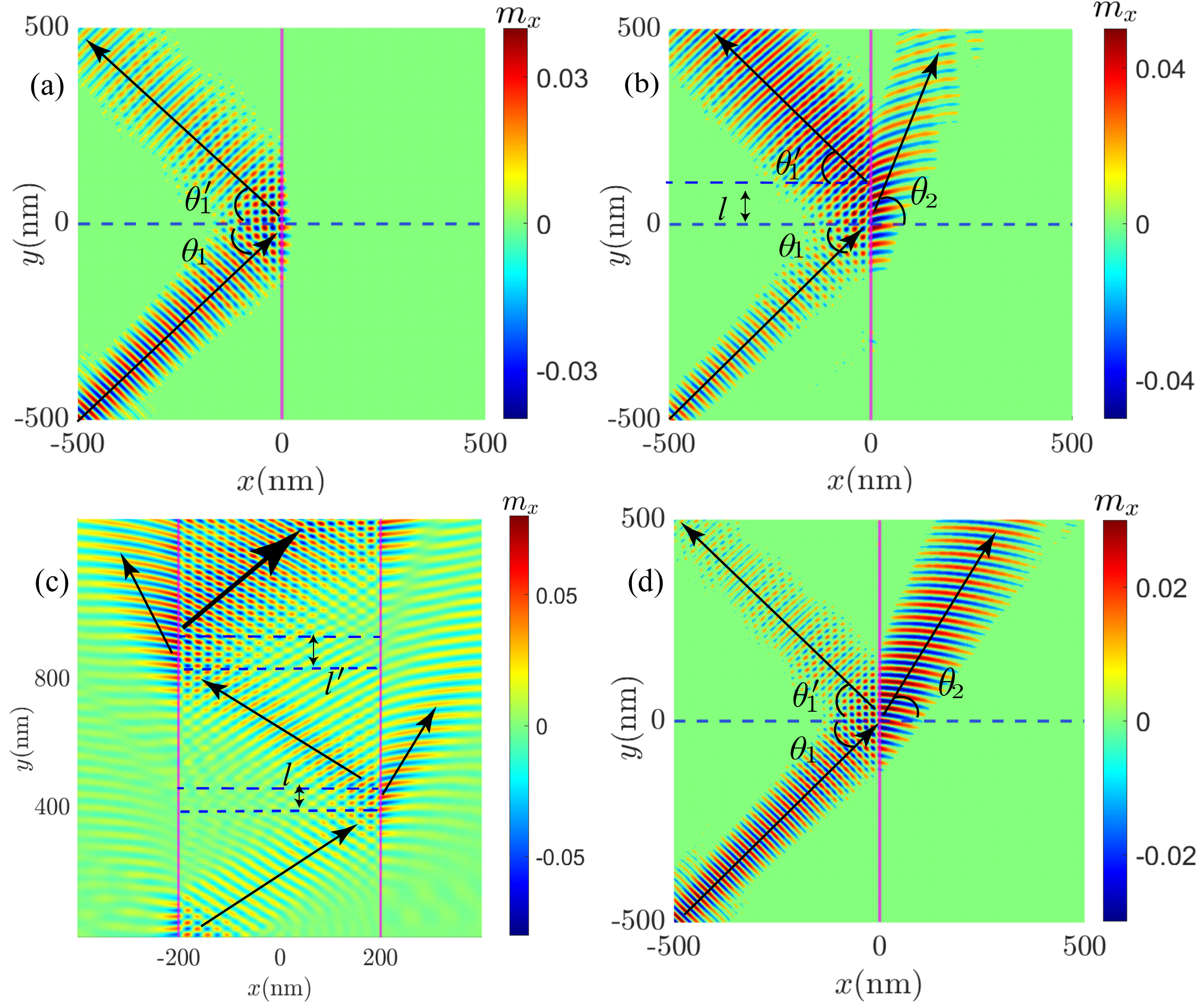} 
\caption{Reflection and refraction of a spin wave of frequency 
$f=\mathrm{100\:GHz}$ and $\alpha=0.001$ at the interface 
without (a) and with (b) a current density of $\vec j = 
\mathrm{10^{13}\hat x \:A\:m^{-2}}$ in $x>0$. The black arrowed 
lines mark the propagation directions of incident, reflective 
and refractive waves. $\theta_{1}=\theta'_{1}=45^{\circ}$, 
$\theta_{2}=71^{\circ}$. 
(c) Spin wave intensity between two super-mirrors (purple lines). 
Along the reflection path (the red arrowed lines) 
the intensity of reflected wave increases. The current density on 
the left (right) region is $j = -\mathrm{10^{13}\hat x
\:A\:m^{-2}}$ ($j = \mathrm{10^{13}\hat x \:A\:m^{-2}}$). 
Wave frequency is $f=\mathrm{105\:GHz}$ and $\alpha=0.001$.
(d) Reflection and refraction of a spin wave of frequency 
$f=\mathrm{120\:GHz}$ and $\alpha=0.001$. 
$\theta_{1}=\theta'_{1}=45^{\circ}$, $\theta_{2}=61^{\circ}$.}
   \label{2D}
\end{figure}
 
Continuity conditions require incident, reflective, and refractive 
wave vectors $\vec k_1, \vec k'_1, \vec k_2$ in general case satisfy 
$k_{1 y}=k'_{1 y}=k_{2 y}$. Denote $\theta_{1}$, $\theta'_{1}$, 
and $\theta_{2}$ incident, reflective, and refractive angles, 
respectively, they follow $\theta'_{1}=\theta_{1}$ and the Snell's 
law \cite{Mulkers2018}, $v_1\sin\theta_1=v_2\sin\theta_2$, where 
1 and 2 stand for normal and pumped regions. 
$v_1$ and $v_2$ are the group velocities \cite{note1}. 
Figure \ref{2D}(a) and (b) show the simulation results for a 
wave of $\theta_{1}=45^{\circ}$ and frequency $f
=\mathrm{100\:GHz}$ , which is in the frequency gap of 
the right region ($x>0$) for $\vec j=0$ and is allowed 
with $\vec j=\mathrm{10^{13}\hat x \:A\:m^{-2}}$.   
Because of the gap, there is no refractive wave when 
$\vec j=0$ as shown in Fig . \ref{2D}(a). The results 
also show $\theta'_{1}=\theta_{1}$ and no amplification 
of the reflected wave since the region is normal for 
$\vec j=0$. In the presence of $\vec j\ne 0$ when the 
wave spectrum is inverted in the right region, Fig. 
\ref{2D}(b) shows clearly the amplification of the 
reflective wave, the presences of both reflective 
and refractive waves with $\theta'_{1}=\theta_{1}$ and 
$v_1\sin\theta_1=v_2\sin\theta_2$ 
(see Supplemental Material for details). 

We further consider a spin wave of $\theta_{1}=45^{\circ}$ 
and $f=\mathrm{105\:GHz}$ propagating in a normal region 
of width $200\,$nm sandwiched between two pumped media with 
$\vec j =-\mathrm{10^{13}\hat x\:A\:m^{-2}}$ ($\vec j=
\mathrm{10^{13}\hat x \:A\:m^{-2}}$) on the left (right) side,
corresponding to spin wave reflection by two super-mirrors. 
Figure \ref{2D}(c) shows clearly that spin waves are amplified 
each times when it is reflected in this flat cavity. 
In contrast to usual laser cavity where waves gain in a 
medium and lose at cavity mirror \cite{Kim2022}, spin 
waves gain at super-mirrors and lose in the medium. 
Thus, the new amplification principle is fundamentally 
different from existing approaches and is an ideal concept 
for microscopic wave amplifiers/lasers.

Reflected waves have lateral shifts of $l$ as shown in Figs. \ref{2D}(b) 
and \ref{2D}(c) when $f$ ($=100, 105\,$GHz) are in the inverted 
region of the pumped medium. Interestingly, such a shift is absent when 
$f$ is in normal region [$\notin (f_{min},f_{max})$] as  
shown in Fig. \ref{2D}(d) for $f=120\,$GHz$>f_{max}$. 
This shift is the Goos-H\"{a}nchen effect or Imbert-Fedorov effect 
\cite{Goos-Hanchen}. In contrast to the shift that normally appears in 
the total reflection in optics for linearly and circularly polarized 
light, we observed here the shift in super-reflection. 
\begin{figure}[h] 
   \centering
   \includegraphics[width=8.5cm]{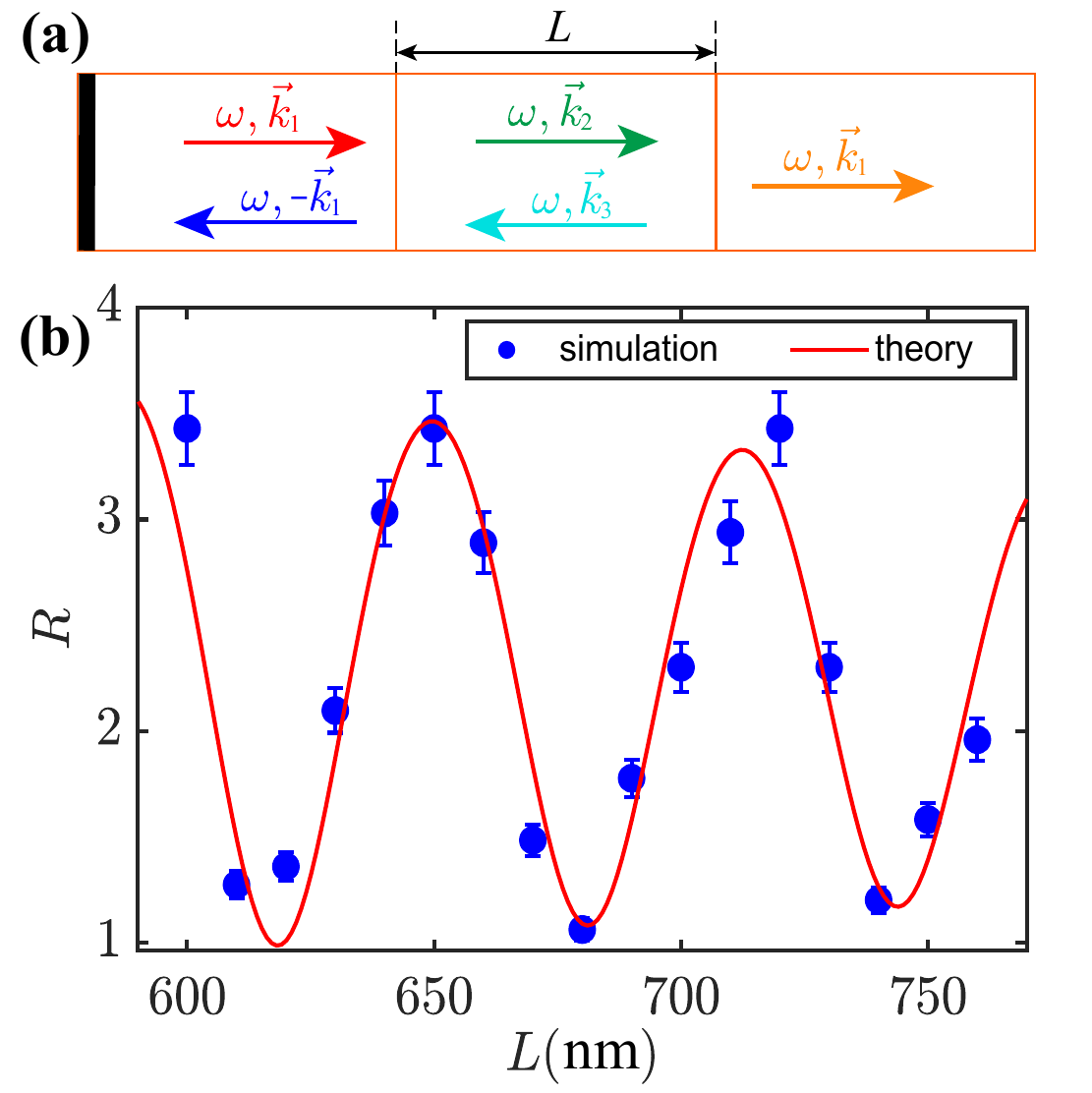} 
   \caption{
(a) A sketch of wave propagation in a sandwiched structure with three regions. Current density of $\vec j=\mathrm{10^{13}\hat x \:A\:m^{-2}}$ only flows in the middle region. 
(b) Reflection coefficient as a function of width $L$ of the middle region. The wave frequency is $f=\mathrm{80\:GHz}$, wave numbers are $k_1 = \mathrm{0.16\:rad/nm}$, $k_2 = \mathrm{-0.03\:rad/nm}$, $k_3 =\mathrm{-0.13\:rad/nm}$, the damping coefficient $\alpha=0.001$, decay constant $\Lambda_2 = \mathrm{0.0008 \,nm}^{-1}$.}
   \label{3domains}
\end{figure}

To gain further insight, we analyse a normally incident wave of $\omega$ 
and $\vec k_1=(k_1,0)$ from left to the right through a pumped region 
of width $L$ sandwiched between two semi-infinite large normal regions. 
We assume that $\omega$ in the inverted spectrum region with corresponding 
two allowed wave vectors $\vec k_2 = (k_2<0, 0)$ and $\vec k_3 = (k_3<0,0)$. 
In another work, incoming wave can activate two backward propagating 
waves in the pumped region. One of them (say $k_2$) has a positive 
group velocity (negative phase velocity because of $k_2<0$), and the 
other ($k_3$) has both negative group and phase velocities. 
The spin wave function in the current case has the form, 
\begin{equation}  
\varphi= \begin{cases}
Be^{-\Lambda_1 x}e^{i(\omega t - k_1 x)}+Ce^{\Lambda_1 x}e^{i(\omega t + k_1 x)}, & x<0 \\
De^{-\Lambda_2 x}e^{i(\omega t - k_2 x)}+Ee^{\Lambda_2 x}e^{i(\omega t - k_3 x)}, & 0\leq x \leq L \\
Fe^{-\Lambda_1 x}e^{i(\omega t - k_1 x)}. & x>L 
\end{cases}
\end{equation}
The wave function should satisfy continuity conditions at both 
$x=0$ and $x=L$. The four conditions determine uniquely reflection 
and transition coefficients, 
\begin{equation}\label{k23}
\begin{aligned}
R =\frac{|C|^2}{|B|^2}  =\left |\frac{(k_1-k_2)D+(k_1-k_3)E}{(k_1+k_2)D+(k_1+k_3)E}\right|^2 \\
\frac{D}{E} =\frac{k_1-k_3}{k_1-k_2}e^{2\Lambda_2 L}e^{i[(k_3-k_2)L+\pi]}
\end{aligned}
\end{equation}
Eq. \ref{reflectioneq}, $R=(\frac{k_1-k_2}{k_1+k_2})^2$, is recovered 
by assuming $E=0$. Eq. \ref{k23} shows that $k_2$ and $k_3$ can interfere 
with each other in the pumped region. The interference is constructive 
when $(k_2-k_3)L=(2n+1)\pi$ and destructive when $(k_2-k_3)L=2n\pi$, 
$n=\mathrm{integer}$. Then the combined backward propagating modes interfere with 
the reflected wave, resulting in a super-reflection $R>1$. 
Figure \ref{3domains} shows how $R$ oscillates with $L$ with a period 
of $2\pi /|k_2-k_3|$ for parameters defined above and $\alpha=0.001$. 

The amplification principle presented here is applicable 
to other waves such as electromagnetic and acoustic waves if similar conditions, such as inverted 
spectrum, an energy source, and proper boundary conditions, are satisfied. 
For spin waves, electric current can be the energy 
source, and it is well-established that current-induced STT can 
invert a spin wave spectrum. 

Compare our spin wave amplification with other existing proposal, 
our proposal is not only universal, but also very easy to implement. Considering the typical spin wave devices are 
very small, super-mirror can be tens nano-meter in size. 
It may also be possible to use several super-mirrors to 
construct a spin wave cavity to make spin-wave ``lasers". 
Similar spin wave amplification by the interface of two coupled
magnets has also been predicted recently \cite{Harms2022}. 
However, the physics in Reference \cite{Harms2022} is 
different from that presented here. Instead of physics of 
constructive interference of the reflective wave with 
incident refractive wave discussed here, Reference
\cite{Harms2022} associates the spin wave amplification from the 
Bosonic Klein paradox that involves the antimagnon, carrying
opposite spin and energy. The spin-orbit torques was 
used to dynamically stabilize both magnons and antimagnons.
In contrast, our theory is based on classical spin wave equations. 
The success of our proposal does not require any quantum effects 
such as quantum tunneling or particle-antiparticle concept and 
the Klein paradox. The spin transfer torques are
used to invert spin wave spectrum. 
We would also like to point out that spin wave spectrum could 
also be inverted with Dzyaloshinskii-Moriya interaction (DMI) 
\cite{Melcher1973, Moon2013}. However, the boundary condition 
of DMI system requires discontinuity in the first derivative of 
spin wave at the boundary \cite{Rohart2013}. This discontinuity 
denied the spin wave amplification discussed here, in agreement 
with the physics principle that spin wave amplification require 
energy source.

In summary, we demonstrate that an interface between a current-free 
region and a current-flow region can be a super-mirror such that 
reflection coefficient can be larger than 1. The physics is the 
refractive wave in a pumped medium is a backward propagating 
wave that can constructively interfere with the reflected wave. 
The effective role of the pumped medium is to coherently emit energy 
to the reflected wave, leading to the superradiance of a super-mirror. 
The proposed physics and picture are verified by the micromagnetic 
simulations.

\begin{acknowledgements}
This work is supported by the National Key Research and Development 
Program of China Grant No. 2020YFA0309600, the National Natural 
Science Foundation of China (Grant No. 11974296) and Hong Kong RGC 
Grants (No. 16301518, No. 16301619, No. 16302321, and No. 16300522).
\end{acknowledgements}


\begin{thebibliography}{99}
\bibitem{note1}In simulations, incoming wave packets around frequency 
$\omega$ and wave vector $\vec k_1=(k_{1x}, k_{1y})$ at the incident 
angle $\theta_1$ is reflected and refracted at the interface. 
The reflected wave packet of frequency $\omega$ and wave vector 
$\vec k'_1=(-k_{1x}, k_{1y})$ travel along its group velocity at the 
reflective angle $\theta'_1=\theta_1$ while the refracted wave packet 
of frequency $\omega$ and wave vector $\vec k_2=(k_{2x}, k_{2y})$ 
travel along its group velocity at refractive angle $\theta_2$. 
In contrast to the Snell law in optics where the phase velocities are 
involved, the Snell law for spin wave packets refraction is determined 
by their group velocities because of their nonlinear wave spectrum.
\end{thebibliography}
\end{document}


\title{Supplemental materials for: Spin wave amplification through superradiance}

\author{X. R. Wang}
\email[Corresponding author: ]{phxwan@ust.hk}
\affiliation{Physics Department, The Hong Kong University of Science and 
Technology, Clear Water Bay, Kowloon, Hong Kong}
\affiliation{HKUST Shenzhen Research Institute, Shenzhen 518057, China}
\affiliation{William Mong Institute of Nano Science and Technology, 
The Hong Kong University of Science and Technology,
Clear Water Bay, Kowloon, Hong Kong, China}

\author{X. Gong}
\affiliation{Physics Department, The Hong Kong University of Science and 
Technology, Clear Water Bay, Kowloon, Hong Kong}

\author{K. Y. Jing}
\affiliation{Physics Department, The Hong Kong University of Science and 
Technology, Clear Water Bay, Kowloon, Hong Kong}

\begin{abstract}
The supplemental materials include  
1) Derivation of spin wave continuity condition at interface
2) Derivation of reflection coefficient of 1D super-mirror and cavity between two super-mirrors, 
3) Derivation of the Snell's law and its numerical verification, 
4) Numerical determinations of spin wave spectrum, reflection and transmission coefficients. 
5) The effect of nonadiabatic STT.
6) An additional micromagnetic simulation of wave amplification in doped permalloy. 

\end{abstract}

\maketitle

\section{Derivation of spin wave continuity condition at interface}
\label{sec1}
$\varphi \equiv m_x + i m_y$ must be continue at interface because $m_x , m_y$
are continuous, or $\varphi(x=0^-) = \varphi(x=0^+)$. Otherwise, the governing 
spin-wave dynamical equation (4) in the main text will contain high order divergent terms. 
To obtain the boundary condition for the derivatives of $\varphi$, we integrate Eq. (4) 
over a volume near the boundary of current-free and current-flow regions. For the simplicity, 
we use 1D as an example. The integral along $x$-direction is  
\begin{equation}
\int_{-\varepsilon}^{\varepsilon}-i\frac{\partial\varphi}{\partial t} dx
=\frac{1+i\alpha}{1+\alpha^2} \int_{-\varepsilon}^{\varepsilon}(-A'\nabla^2+K'+i\vec{u}\cdot\nabla)\varphi dx .
\end{equation}
As $\varepsilon \to 0$, the term on left-hand side
\begin{equation}
\lim_{\varepsilon \to 0}\int_{-\varepsilon}^{\varepsilon}\frac{\partial\varphi}{\partial t} dx = 0,
\end{equation}
the first and second term on right-hand side
\begin{equation}
\lim_{\varepsilon \to 0}\int_{-\varepsilon}^{\varepsilon} A' \nabla^2\varphi dx 
= \lim_{\varepsilon \to 0}A' \frac{\partial\varphi}{\partial x}\bigg|_{-\varepsilon}^{\varepsilon}
= A' \left[\frac{\partial\varphi}{\partial x}(x=0^+) - \frac{\partial\varphi}{\partial x}(x=0^-)\right],
\end{equation}
\begin{equation}
\lim_{\varepsilon \to 0}\int_{-\varepsilon}^{\varepsilon} K' \varphi dx 
= \lim_{\varepsilon \to 0} \left(K'(x<0)\int_{-\varepsilon}^{0}  \varphi dx + 
K'(x\geq0) \int_{0}^{\varepsilon} \varphi dx \right) = 0,
\end{equation}
Integrating the third term on right-hand side by parts gives
\begin{equation}
\lim_{\varepsilon \to 0}\int_{-\varepsilon}^{\varepsilon} (\vec{u}\cdot\nabla)\varphi dx 
= \lim_{\varepsilon \to 0} u\varphi\bigg|_{-\varepsilon}^{\varepsilon} 
- \lim_{\varepsilon \to 0}\int_{-\varepsilon}^{\varepsilon}  \varphi \nabla\cdot\vec{u}dx.
\end{equation}
Since $\vec{u} = 0 (x<0)$, $\vec{u}=u\hat{x}(x\geq0)$, $\frac{\partial\vec{u}}{\partial x}= u\delta(x)\hat{x}$,
\begin{equation}
\lim_{\varepsilon \to 0} u\varphi\bigg|_{-\varepsilon}^{\varepsilon} 
= \lim_{\varepsilon \to 0} u(\varepsilon)\varphi(\varepsilon)-u(-\varepsilon)\varphi(-\varepsilon)
= \lim_{\varepsilon \to 0} u(\varepsilon)\varphi(\varepsilon) - 0 = u\varphi(x=0),
\end{equation}
and
\begin{equation}
\lim_{\varepsilon \to 0}\int_{-\varepsilon}^{\varepsilon}  \varphi \nabla\cdot\vec{u}dx
= \lim_{\varepsilon \to 0}\int_{-\varepsilon}^{\varepsilon}  \varphi u \delta(x)dx = u \varphi(x=0).
\end{equation}

There is another way to see $\nabla\varphi(0^-) = \nabla\varphi(0^+)$ directly from Eq. (4). 
Since all terms, except possibly the first term on the right hand side, in the 
equation are finite (the last term on the right hand side can at most be a step function), the tiny 
integrals around boundary must be order of infinitesimal volume. 
To the zero order in the small volume, the integral is 
$A[\nabla\varphi(0^+)-\nabla\varphi(0^-)]=0$, which is the claimed boundary condition.

\section{Derivation of reflection coefficient of 1D super-mirror and cavity between two super-mirrors}
\label{sec2}
For a wave propagating from the left to the right along the $x$-direction 
in one dimensional space with an interface at $x=0$ which separates two 
homogeneous media with different material parameters, the wave function 
of frequency $\omega$ is, due to the reflection and refraction,   
\begin{equation}  
\varphi= \begin{cases}
Be^{-\Lambda_1 x}e^{i(\omega t - k_1 x)}
+Ce^{\Lambda_1 x}e^{i(\omega t + k_1 x)}, & x<0 \\
De^{-\Lambda_2 x}e^{i(\omega t - k_2 x)}, & x\geq0 . \
\end{cases}
\label{wavefunction}
\end{equation}
$k_1$ and $-k_1$ are two corresponding wave vectors of $\omega$ in medium 
1 on the left, corresponding to the forward and backward propagating waves. 
$k_2$ is one of two wave vectors of $\omega$ in medium 2 on the right 
side of the interface whose group velocity is toward the right. 
$\Lambda_i^{-1}$ ($i=1,2$) is the decay length in medium $i$.
$\varphi$ satisfies the continuity conditions at the interface ($x=0$), 
$\varphi(x=0^-)=\varphi(x=0^+)$ and
$\nabla\varphi(x=0^-)=\nabla\varphi(x=0^+)$. 
Thus, 
\begin{equation}
\begin{aligned}
B+C&=D,\\
(-ik_1-\Lambda_1)B+(ik_1+\Lambda_1)C&=(-ik_2-\Lambda_2)D.
\end{aligned}
\end{equation}
$\frac{C}{B} = \frac{(k_1-i\Lambda_1)-(k_2-i\Lambda_2)}{(k_1-i\Lambda_1)+(k_2-i\Lambda_2)}$,
$\frac{D}{B} = \frac{2(k_1-i\Lambda_1)}{(k_1-i\Lambda_1)+(k_2-i\Lambda_2)}$. 
When the decay lengths are much larger than the wavelength, the reflection and 
transmission coefficients are
\begin{equation}
\begin{aligned}
R=\frac{|C|^2}{|B|^2}=\left (\frac{k_1-k_2}{k_1+k_2}\right )^2 \\
T=\frac{k_2 |D|^2}{k_1 |B|^2}=\frac{4 k_1 k_2}{(k_1+k_2)^2}.
\end{aligned}
\end{equation}

In the case of a wave propagating from the left to the right through 
two super-mirrors at $x=0$ and $x=L$, respectively, which consists of 
a pumped medium in the middle ($0\leq x\leq L$) and two identical normal 
media on the two sides, $\varphi$ is,  
\begin{equation}  
\varphi= \begin{cases}
Be^{-\Lambda_1 x}e^{i(\omega t - k_1 x)}+Ce^{\Lambda_1 x}e^{i(\omega t + k_1 x)}, & x<0 \\
De^{-\Lambda_2 x}e^{i(\omega t - k_2 x)}+Ee^{\Lambda_2 x}e^{i(\omega t - k_3 x)}, & 0\leq x \leq L \\
Fe^{-\Lambda_1 x}e^{i(\omega t - k_1 x)}, & x>L 
\end{cases}
\end{equation}
where $k_2$ and $k_3$ are assumed to be two negative wave vectors in the pumped media. 
The spectrum has a positive group velocity at $k_2$ and a negative group velocity at $k_3$.
The boundary conditions yield  
\begin{equation}\label{3region}
\begin{aligned}
&B+C=D+E\\
&(k_1-i\Lambda_1) B - (k_1-i\Lambda_1) C = (k_2-i\Lambda_2) D + (k_3+i\Lambda_2) E\\
&De^{-\Lambda_2 L}e^{-ik_2L}+Ee^{\Lambda_2 L}e^{-ik_3L}=Fe^{-\Lambda_1 L}e^{-ik_1L}\\
&(k_2-i\Lambda_2) De^{-\Lambda_2 L}e^{-ik_2L}+ (k_3+i\Lambda_2)Ee^{\Lambda_2 L}e^{-ik_3L}
= (k_1-i\Lambda_1) Fe^{-\Lambda_1 L}e^{-ik_1L}. \\
\end{aligned}
\end{equation}
In terms of $k_1, k_2, k_3\gg \Lambda_1, \Lambda_2$, we have 
\begin{equation}
\begin{aligned}
&R =\frac{|C|^2}{|B|^2}  =\left |\frac{(k_1-k_2)D+(k_1-k_3)E}{(k_1+k_2)D+(k_1+k_3)E}\right|^2 \\
&\frac{D}{E} =-\frac{k_1-k_3}{k_1-k_2}e^{2\Lambda_2 L}e^{i(k_3-k_2)L}.
\end{aligned}
\end{equation}

\section{Derivation of the Snell's law and its numerical verification}
\label{sec3}
For a spin wave of frequency $\omega$ with incident angle $\theta_1$ in a 
two-dimensional plane, the wave function is  
\begin{equation}  
\varphi= \begin{cases}
Be^{-\Lambda_1 x}e^{i(\omega t - \vec{k}_1 \cdot\vec{r})}
+Ce^{\Lambda_1 x}e^{i(\omega t - \vec{k}'_1 \cdot\vec{r})}, & x<0 \\
De^{-\Lambda_2 x}e^{i(\omega t - \vec{k}_2 \cdot\vec{r})}, & x\ge0 \\
\end{cases}
\end{equation}
where $|\vec{k}_1|=|\vec{k}'_1|$ and $k_{1,x}=-k'_{1,x}$ for spectrum 
in current-free region. $\vec{k}_2$ is the wave vector of frequency 
$\omega$ whose group velocity is forward (positive). From the boundary 
conditions of $\varphi(x=0^-)=\varphi(x=0^+)$ and $\nabla\varphi
(x=0^-)=\nabla\varphi(x=0^+)$, we have 
\begin{equation}
\begin{aligned}
Be^{-i {k}_{1,y} y}+Ce^{ -i {k}'_{1,y} y }
&=De^{ -i {k}_{2,y} y} \\
\vec{k}_1Be^{ -i {k}_{1,y} y }+\vec{k}'_1 Ce^{ -i {k}'_{1,y}y}
&=\vec{k}_2 De^{ -i{k}_{2,y} y}.
\end{aligned}
\label{BC}
\end{equation}
Above equations are true for an arbitrary $y$ only when ${k}_{1,y}=
k'_{1,y}={k}_{2,y}$. Here we are dealing with a wave packet centred 
around frequency $\omega$ which is emitted from a source located at the 
lower left corner of the sample as shown in Fig. 1 in the main text. 
The wave-packet propagates along group velocity of $\vec{v}_2=2A'\vec{k}_2+
\vec{u}$ in the pumped region while the incident and reflective waves propagate 
along $\vec{v}_1=2A'\vec{k}_1$ and $\vec{v}'_1=2A'\vec{k}'_1$, respectively. 
If $\theta_1$, $\theta'_1$, and $\theta_2$ are defined as incident, reflective, 
and refractive angles (the angle between group velocity and $\hat{x}$),
then we have $v_{1, y} = v'_{1, y} = v_{2, y}$, $\theta_1=\theta'_1$, and 
\begin{equation}
v_1\sin\theta_1=v_2\sin\theta_2. 
\label{snell}
\end{equation}

To verify the correctness of the Snell's law, we consider a film of 
$2000\times2000\times2 \mathrm{nm}^3$ centred at the origin of $xy$-plane. 
All model parameters are the same as those in the main text, and the left-half 
of the film is current-free while the right-half of the film has current 
flowing to the $x$-direction. We use MuMax3 to solve the LLG equation with 
an oscillating magnetic field of frequency $\omega$ applied on a strip of 
$100\mathrm{nm}\times4\mathrm{nm}$ which is placed at the lower-left 
corner of the film at various inclined angles with respect to $y$-axis.
In order to eliminate un-wanted spin wave reflection from the film 
boundary, the damping coefficient at the boundaries is assumed to be 0.5.
The generated spin wave propagates toward the interface at an arbitrary 
incident angle $\theta_1$ that is determined by the inclined angle of 
the strip. After the magnetization reaches its stable distribution, 
$\vec m(\vec x) \cos[\omega t-\phi(\vec x)]$, we use those $|m_x(x, y)|
>0.2\max(|m_x(x, y)|)$ or $|m_y(x, y)|>0.2\max(|m_y(x, y)|)$ to compute the 
incident, reflective and refractive spin wave beams. We then numerically 
search $y$'s such that $|\vec m(x,y)|$ is the local maxima for a given $x$. 
The beam directions are defined as the lines connecting all these local 
maximal field points. The incident, reflective and refractive angles 
can then easily be obtained from the beam directions. 
Wave vector $\vec{k}=(k_x, k_y)$ is obtained through the two-dimensional 
Fourier transformation $m_x(x, y) \to \tilde{m}_x(k_x, k_y)$ (the peak 
position of $|\tilde{m}_x(k_x, k_y)|^2$ is defined as $k_x$ and $k_y$).
The group velocity is then obtained from the formula $v=\sqrt{(2A'k_x)^2
+(2A'k_y)^2}$ (current-free region) or $\sqrt{(2A'k_x+u)^2+(2A'k_y)^2}$ 
(current-flow region). The results are listed in Table. \ref{angle}, 
which follows the Snell's law (Eq. \ref{snell}).
\begin{table}[H]
\caption{Incident angles $\theta_1$, refractive angles $\theta_2$, 
group velocities in medium 1 ($v_1$) and in medium 2 ($v_2$)}
\begin{center}
\begin{tabular}{|c|c|c|c|c|c|c|}
\hline
Frequency $f$(GHz) & Incident angle $\theta_1$ & Refractive angle $\theta_2$ & $v_{1}$(m/s) & $v_{2}$(m/s) & $v_1\sin\theta_1$(m/s) & $v_2\sin\theta_2$(m/s)\\ 
\hline
90 & $30^{\circ}$  & $57^{\circ}$ & 2239 & 1348 & 1120 & 1130 \\
\hline
100 & $45^{\circ}$  & $71^{\circ}$ & 2544 & 1881 & 1799  & 1778 \\
\hline
110 & $45^{\circ}$  & $65^{\circ}$ & 2821 & 2183 & 1995  & 1978 \\
\hline
120 & $45^{\circ}$  & $61^{\circ}$ & 3072 & 2498 & 2172  & 2184 \\
\hline
\end{tabular}
\end{center}
\label{angle}
\end{table}

\section{Numerical determinations of spin wave spectrum, reflection and transmission coefficients}
\label{sec4}
Spin wave spectrum of a given system can be extracted from a thermally activated 
magnetization fluctuation around its equilibrium state from the Fourier transformation 
of the fluctuation as used in our previous studies \cite{Xiansi1,Xiansi2,Ying1,Ying2}.
We use a magnetic strip of $2000\times2\times2 \mathrm{nm}^3$ as an example to illustrate 
the procedures. The material parameters are the same as those given in the main text.  
At a finite temperature, magnetization can feel a thermal field of $\vec{h}=\vec{\eta}
\sqrt{2\alpha k_{\rm B} T/(M_{\rm s} \mu_0 \gamma\Delta V\Delta t)}$, where $\Delta V$, 
$\Delta t$, $T$ and $\vec{\eta}$ are the cell volume, time step, temperature, and a 
random vector from a standard normal distribution, respectively \cite{mumax3,Brown1963}. 
MuMax3 is used to generate magnetization fluctuation around its thermal equilibrium state. 
Starting from an initial magnetization state of $\vec{m}=\hat{z}$, magnetization 
will soon (in nanoseconds) reaches its thermal equilibrium states. 
$m_x(x, t)$ and  $m_y(x, t)$ are then recorded every $0.5\mathrm{ps}$ for $1 \mathrm{ns}$. 
This shall allow us to find spin wave spectrum for frequencies between $f =1$GHz
to $f=2$THz. From the Fourier transform $\tilde\varphi(k, f)$ of 
$\varphi (x, t)\equiv m_x(x, t)+im_y(x, t)$, the spectral intensity function is then 
the square of its magnitude, $|\tilde\varphi(k, f)|^2$. 
The spin wave spectrum can then easily be obtained from the density 
plot of $|\tilde\varphi(k, f)|^2$ as a color map in $fk$-plane, as shown in Fig. 2(c) and (d) in the main text.  

\begin{figure}
	\centering
	\includegraphics[width=15cm]{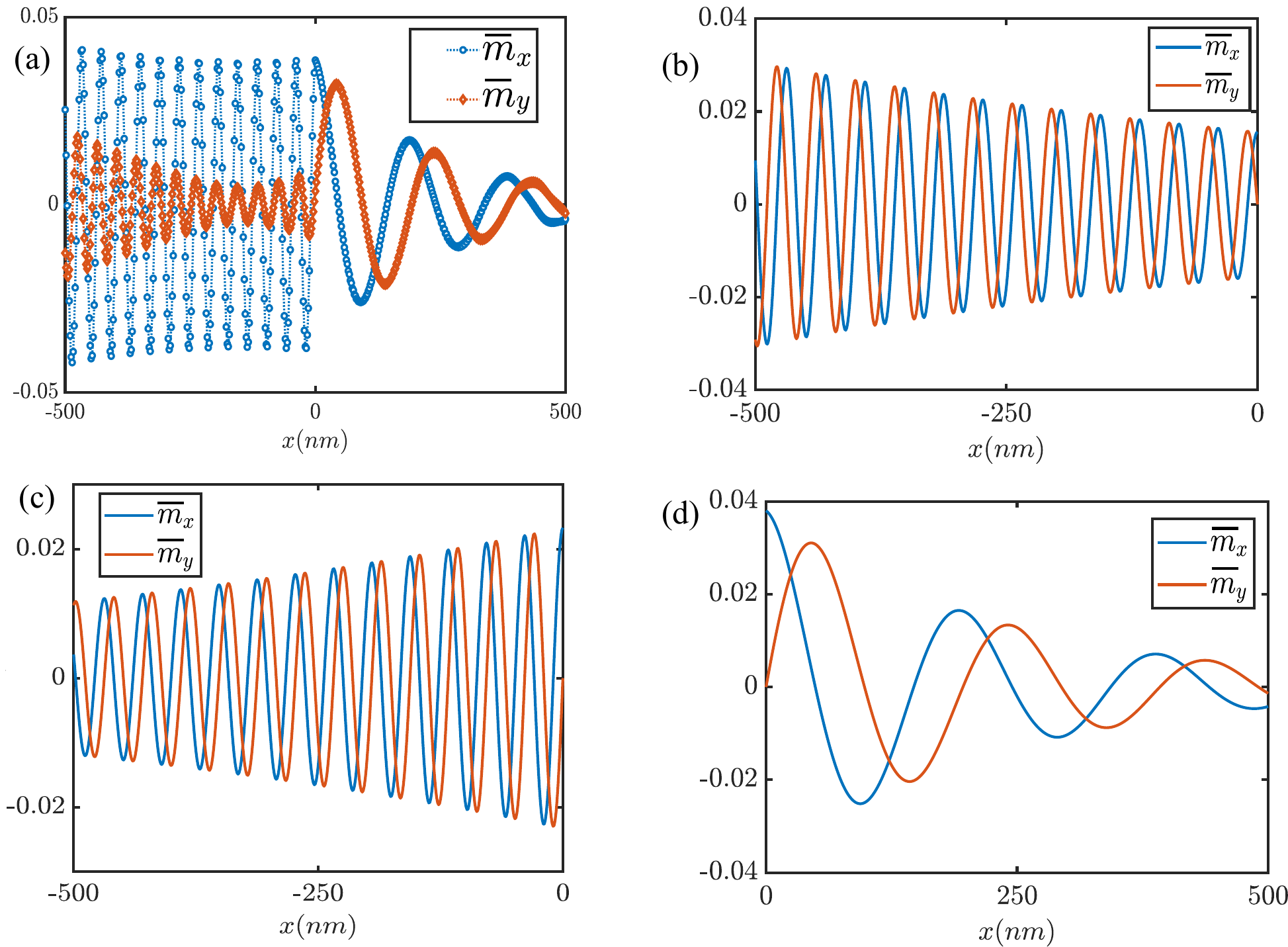} 
\caption{(a) A snap-shot of spin wave of $\bar{m}_x\equiv \frac{1}{16}\sum_y 
m_x(x, y)$ and $\bar{m}_y\equiv\frac{1}{16}\sum_y m_y(x, y)$ at $t=2$ns. 
$f=80$GHz  is used. The interface between the current-free region 
(left) and current-flow region (right) is at $x=0$.
(b) The incident wave. (c) The reflective wave. (d) The refractive wave.}
\label{wave}
\end{figure}

\begin{figure}
	\centering
	\includegraphics[width=15cm]{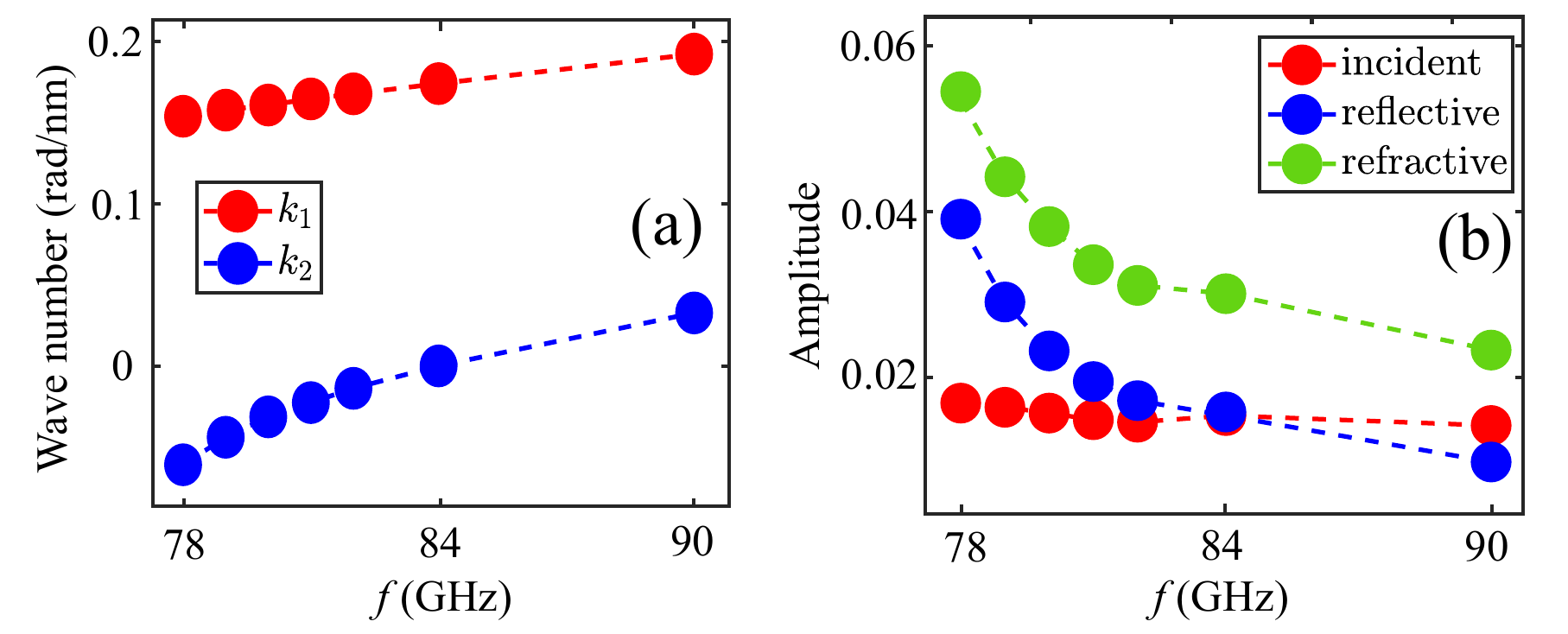} 
\caption{(a) Wave numbers $k_1$ (red dots) and $k_2$ (blue dots) 
dependence on frequency obtained from simulations. 
(b) Amplitudes of the incident wave (red dots), reflective wave (blue dots) , 
and refractive wave (green dots) for spin wave with different frequencies.}
\label{data}
\end{figure}

To simulate reflection and refraction of a normally incident spin wave, an 
oscillating magnetic field $\mu_0 \vec{H}=0.5\sin(\omega t)\hat{x}\mathrm{T}$ 
is applied in a $32\times 4\mathrm{nm}^2$ strip on the left boundary of the 
current-free region. The film size is $2000\times32\times2 \mathrm{nm}^3$. 
We choose $f$ in the range of $[78, 90]\: \mathrm{GHz}$, which covers 
both the frequency gap and inverted frequency in the pumped region.
When the system reaches a dynamic equilibrium, spin wave $\varphi(x, t)=
\frac{1}{16}\sum_y m_x(x, y, t)+i m_y(x, y, t)$ is then obtained.
Then we use the Prony method \cite{prony} (fitting the data by wave function 
Eq. \ref{wavefunction}) to separate the incident, reflective and refractive 
waves. The results are shown in Fig. \ref{wave}. The Prony method allows 
us to extract $|B|$, $|C|$, $|D|$, $k_1$, $k_2$, $\Lambda_1^{-1}$, and 
$\Lambda_2^{-1}$ which are plotted in Fig. \ref{data}. 
The reflection and transmission coefficients defined as $R=|C|^2/|B|^2$ and 
$T = k_2|D|^2/(k_1|B|^2)$ are given in Fig. 3 in the main text. 

\section{The effect of non-adiabatic STT}
\label{sec5}
In this section, we add a non-adiabatic STT, 
$\beta \vec{m}\times (\vec{u}\cdot\nabla)\vec{m}$, to the LLG equation,
and investigate how it affects the spin wave amplification.
The dynamic equation of $m_x$ and $m_y$ in current-flow region becomes 
\begin{equation}
\left(\begin{array}{cc}
\partial_t+\vec{u}\cdot\nabla & -A^{\prime} \nabla^2+K^{\prime}+\alpha \partial_t+\beta \vec{u}\cdot\nabla \\
A^{\prime}\nabla^2 - K^{\prime}-\alpha \partial_t-\beta \vec{u}\cdot\nabla & \partial_t+\vec{u}\cdot\nabla
\end{array}\right)\left(\begin{array}{l}
m_x \\
m_y
\end{array}\right)=0 .
\end{equation}
Then the dispersion relation is 
\begin{equation}
\omega - i \alpha \omega = A' |\vec{k} - i\vec{\Lambda}|^2 + \vec{u}\cdot (\vec{k}
- i \vec{\Lambda})  + K' - i\beta \vec{u}\cdot (\vec{k} - i \vec{\Lambda}) . 
\end{equation}
Taking one-dimensional case as an example,  the equation becomes 
\begin{subequations}
\begin{equation}
A' \left(k + \frac{u}{2A'}\right)^2   = \omega - K' + \frac{u^2}{4A'}+ A' \Lambda^2 +\beta u \Lambda ,
\end{equation}
\begin{equation}
 (2 A' k + u) \Lambda  = \alpha \omega -  \beta u k . 
\end{equation}
\label{lambdak}
\end{subequations}
For a given $\omega$, the corresponding wave number of refractive wave takes the solution 
\begin{equation}\label{betakw}
\begin{split}
k_2 - i\Lambda_2  
&= \frac{-(u-i\beta u)+\sqrt{(u-i\beta u)^2+4A'(\omega - i\alpha \omega - K'_2)}}{2A'} \\
&= \frac{-(u-i\beta u)+\sqrt{u^2 -\beta^2 u^2 +4A' (\omega - K'_2) - i (2\beta u^2 + 4A'\alpha \omega)}}{2A'}\\
& = \frac{u}{2A'} \left[ -1+i\beta +\sqrt{1 -\beta^2  +4A' (\omega - K'_2)/u^2 - i (2\beta  + 4A'\alpha \omega/u^2)} \right] .
\end{split}
\end{equation} 
Define $a = 1 -\beta^2 +4A' (\omega - K'_2)/u^2 $,  $b = 2\beta  + 4A'\alpha \omega/u^2$.
In most realistic cases \cite{Thomas2006,Hayashi2006,Moriya2008,Heyne2008, Liu2020, Jin2021, Wang2023},
$\alpha, \beta \ll 1$, $b/a \ll 1$, then we can expand the square root term to the 
second order of $b/a$: 
\begin{equation}
\sqrt{1 -\beta^2  +4A' (\omega - K'_2)/u^2 - i (2\beta  + 4A'\alpha \omega/u^2)}= \sqrt{a - bi} \approx \sqrt{a}\left(1 - \frac{bi}{2a} + \frac{b^2}{8a^2}\right).
\end{equation}
Thus, 
\begin{equation}
\Lambda_2 = \frac{u}{2A'} \left(-\beta  +\sqrt{a}\frac{b}{2a}\right)  = \frac{u}{2A'}\left[\frac{2A'\alpha\omega}{u^2\sqrt{a}} 
+ \beta  \left(\frac{1}{\sqrt{a}}-1 \right) \right]. 
\end{equation}
Under the condition of superradiance,  $k_2 <0$, $\omega < K'_2$,  $a < 1$, the 
term proportional to $\beta$ in the square bracket is positive, indicating an increase of decay constant $\Lambda_2$ from non-adiabatic STT. 
Further, the increase of $\Lambda_2$ leads to a subsequent rise in $ A'{\Lambda_2}^2 + \beta u \Lambda_2$, 
resulting in an increase of $k_2$ (less negative), as seen from Eq. \ref{lambdak}(a). 
For the wave number, 
\begin{equation}
k_2 = \frac{u}{2A'}\left(-1 + \sqrt{a} + \sqrt{a}\frac{b^2}{8a^2}\right) .
\end{equation}
Using relations
\begin{equation}
\left(\sqrt{a} + \sqrt{a}\frac{b^2}{8a^2}\right)^2 > a + \frac{b^2}{4a} , 
\end{equation}
and
\begin{equation}
 \frac{b^2}{4a}=\frac{(2\beta  + 4A'\alpha \omega/u^2)^2 /4}
 {1 -\beta^2 +4A' (\omega - K'_2)/u^2}
> \frac{\beta^2}{1 -\beta^2 +4A' (\omega - K'_2)/u^2}, 
\end{equation}
one has 
\begin{equation}
\left(\sqrt{a} + \sqrt{a}\frac{b^2}{8a^2}\right)^2 
>1 -\beta^2 +4A' (\omega - K'_2)/u^2 + \frac{\beta^2}{1 -\beta^2 +4A' (\omega - K'_2)/u^2}=1 +4A' (\omega - K'_2)/u^2 + \beta^2\left(\frac{1}{a} -1\right)>1 +4A' (\omega - K'_2)/u^2, 
\end{equation}
and 
\begin{equation}
-1+\sqrt{a} + \sqrt{a}\frac{b^2}{8a^2}
>-1+\sqrt{1 +4A' (\omega - K'_2)/u^2}.
\end{equation}
Thus,  
\begin{equation}
k_2  > \frac{u}{2A'}\left[-1 + \sqrt{1 +4A' (\omega - K'_2)/u^2}  \right ],
\end{equation}
value of $k_2$ in the presence of $\beta$ is larger (less negative) than that in the absence of $\beta$.
We can conclude that the presence of the non-adiabatic STT
leads to an increase of wave number. 

Since $\Lambda_2$ is not negligible, the exact reflection coefficient should be 
\begin{equation}\label{Rbeta}
R=\left |\frac{k_1-k_2+i\Lambda_2}{k_1+k_2 - i\Lambda_2}\right |^2  = \frac{(k_1 - k_2)^2 + {\Lambda_2}^2}{(k_1 + k_2)^2 + {\Lambda_2}^2} . 
\end{equation}
When $k_1 >0$, $k_2 < 0$, the increasing of $\Lambda_2$ and $k_2$ 
will lead to the reduction of $R$.

\begin{figure}[H]
	\centering
	\includegraphics[width=18cm]{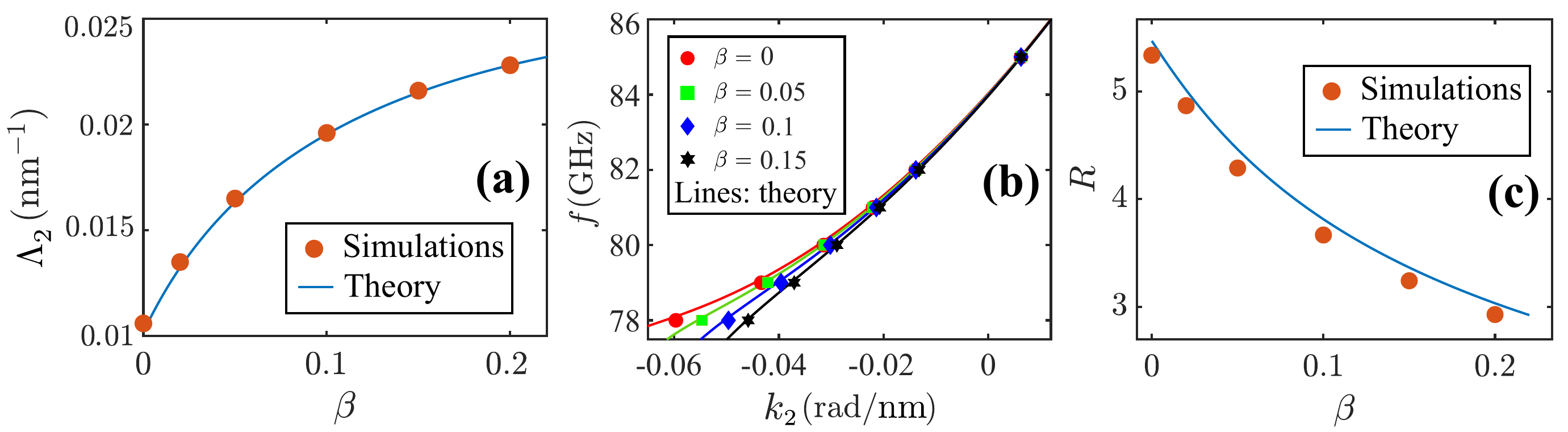} 
\caption{(a) {$\beta$-dependence of $\Lambda_2$ for spin wave  frequency $f=78$ GHz. 
(b) Spin wave spectrum ($f-k_2$ relations)} for different $\beta$. 
For a given frequency, $k_2$ increases as $\beta$ increases.
(c) $\beta$-dependence of $R$ for $f=78$ GHz. 
$\alpha = 0.005$ and various $\beta$ in the range of $[0,40\alpha]$
is used throughout the simulations. }
\label{beta}
\end{figure}

These analysis (solid lines in Fig. \ref{beta}) are verified by micromagnetic simulations (dots in Fig. \ref{beta}).
From Fig. \ref{beta} (a,b), we can see that for a given frequency, the $\Lambda_2$ and $k_2$ increase with increasing $\beta$, and reflection coefficient decreases with the increase of $\beta$, in agreement with our theoretical analysis not only 
qualitatively, but also quantitatively.
The range of $\beta$ is taken as $[0,40\alpha]$, as in most common materials 
\cite{Thomas2006,Hayashi2006,Moriya2008,Heyne2008, Liu2020, Jin2021, Wang2023}. 
For extremely large $\beta$ above this range, dynamical 
instabilities may occur.

\section{Simulation results in doped permalloy}
\label{sec5}

In our theory, the frequency window width in which spin waves can be amplified is $f_{\mathrm{max}}-f_{\mathrm{min}}=\frac{1}{2\pi}\frac{u^2}{4A'}= \frac{P^2 \mu_0 \mu_B^2  j^2}{16\pi \gamma e^2 M_s A}$.
For material (Co/Pt) parameters used in main text, the required 
current density for spin wave amplification at useful frequencies is order of $10^{13}\:\mathrm{A}\:\mathrm{m}^{-2}$. 
One may use point contact structures \cite{Tsoi1998, Ji2003} 
to realized such a high current density and corresponding 
frequency window width is about 6 GHz. 
Material with lower saturation magnetization $M_s$ and exchange stiffness $A$, the required current density can be lower.   
For silver (Ag)-doped permalloy (Py) \cite{Yin2015} with  
$M_s = \mathrm{0.39\sim0.7 \:MA\:m^{-1}}$ and $A = \mathrm{2.8\sim 11 \:pJ\:m^{-1}}$, 
the required current density can be reduced to the order of 
$10^{12} \:\mathrm{A}\:\mathrm{m}^{-2}$ 
(See e.g. Ref. \cite{Meo2022} for possible experimental realization) with a frequency window width of 0.5 GHz. 

\begin{figure}[H]
	\centering
	\includegraphics[width=18cm]{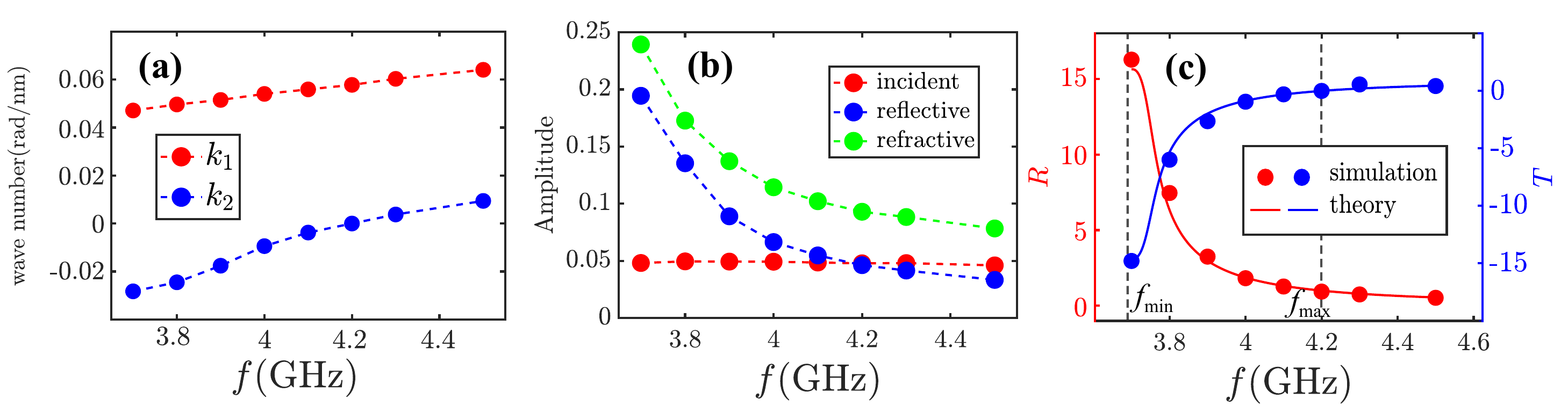} 
\caption{ (a) $k_1$-dependence (red dots) and $k_2$-dependence  (blue dots) of frequency. 
(b) Amplitudes of the incident wave (red dots), reflective wave (blue dots) , 
and refractive wave (green dots) as functions of frequency.
(c) Reflection $R$ and transmission coefficient $T$ versus frequency $f$. The red and blue solid lines are theoretical predictions of Eq. \eqref{Rbeta}. 
Two dashed lines mark $f_{\mathrm{min}}$ and 
$f_{\mathrm{max}}$, respectively. }
\label{permalloy}
\end{figure}

In our simulations showed below, the applied current density is 
$j = 1.2 \times 10^{12}\:\mathrm{A}\:\mathrm{m}^{-2}$ with $P=1$, 
$M_s = \mathrm{0.4 \:MA\:m^{-1}}$, $A = \mathrm{3 \:pJ\:m^{-1}}$,
$K_1 =0.02 \:\mathrm{MJ}\:\mathrm{m}^{-3}$, 
$K_2=0.03 \:\mathrm{MJ}\:\mathrm{m}^{-3}$, non-adiabatic STT 
coefficient $\beta = 0.01$, and $\alpha = 0.005$.
The frequency of spin wave with zero wave number is $2.8 \,\mathrm{GHz}$ in current-free region  
and $4.2 \,\mathrm{GHz}$ in current-flow region. 
The frequency window of amplification is 
$f \in [3.7, 4.2]\,\mathrm{GHz}$. 
The simulation procedures are the same as those in Sec. \ref{sec4}. 
The wave numbers and
amplitudes extracted by using Prony method are 
presented in Fig. \ref{permalloy} (a) and (b).
Fig \ref{permalloy} (c) shows clearly supper-reflection $R>1$ 
($T<0$) in at frequency $f\in [f_\mathrm{min}, f_\mathrm{max}] $ and $k_2 <0$.
